\documentclass[aip,jvst,reprint,frontmatterverbose]{revtex4-1}
\usepackage{graphicx}
\usepackage{ulem}
\usepackage{amssymb}
\usepackage{amsmath}
\usepackage{todonotes}
\usepackage{wasysym}
\usepackage{relsize}
\usepackage{bm}
\bibpunct{}{}{,}{s}{}{}
\usepackage{float}
\usepackage{epstopdf}

\begin{document}

\title{Interpreting the field emission equation for large area field emitters}

\author{Debabrata Biswas}\email{dbiswas@barc.gov.in}
\affiliation{
Bhabha Atomic Research Centre,
Mumbai 400 085, INDIA}
\affiliation{
  Homi Bhabha National Institute, Mumbai 400 094, INDIA}

\begin{abstract}
  Both single emitters and large area field emitters (LAFE) are generally characterized using
  the slope and intercept of a Murphy-Good (or Fowler-Nordheim)
  plot which are used to extract the field enhancement factor and the emission area.
  Using a shielding model that has been developed recently for a LAFE,
  the validity of the underlying assumption is investigated.
  It is found that in case of a LAFE, the slope has contributions from
  the enhancement factor {\it as well as} the rate at which the effective number of
  super-emitters changes with the applied field.
  As a consequence, the emission area is related to both the
  slope and the intercept in a LAFE. When the mean spacing in a LAFE is much larger than
  the height of emitter, the usual interpretation of the slope and intercept are recovered. 
\end{abstract}

\maketitle

\newcommand{\be}{\begin{equation}}
\newcommand{\ee}{\end{equation}}
\newcommand{\bea}{\begin{eqnarray}}
\newcommand{\eea}{\end{eqnarray}}
\newcommand{\Tbar}{{\bar{T}}}
\newcommand{\En}{{\cal E}}
\newcommand{\K}{{\cal K}}
\newcommand{\GC}{{\cal \tt G}}
\newcommand{\Lop}{{\cal L}}
\newcommand{\DB}[1]{\marginpar{\footnotesize DB: #1}}
\newcommand{\q}{\vec{q}}
\newcommand{\kt}{\tilde{k}}
\newcommand{\Lopn}{\tilde{\Lop}}
\newcommand{\noi}{\noindent}
\newcommand{\ovn}{\bar{n}}
\newcommand{\ovx}{\bar{x}}
\newcommand{\ovE}{\bar{E}}
\newcommand{\ovV}{\bar{V}}
\newcommand{\ovU}{\bar{U}}
\newcommand{\ovJ}{\bar{J}}
\newcommand{\calE}{{\cal E}}
\newcommand{\ovphi}{\bar{\phi}}
\newcommand{\zt}{\tilde{z}}
\newcommand{\rt}{\tilde{\rho}}
\newcommand{\tth}{\tilde{\theta}}
\newcommand{\nuv}{{\rm v}}
\newcommand{\ck}{{\cal K}}
\newcommand{\cc}{{\cal C}}
\newcommand{\ca}{{\cal A}}
\newcommand{\cb}{{\cal B}}
\newcommand{\cg}{{\cal G}}
\newcommand{\ce}{{\cal E}}
\newcommand{\cn}{{\cal N}}
\newcommand{\fn}{{\small {\rm  FN}}}
\newcommand\norm[1]{\left\lVert#1\right\rVert}
\newcommand{\Afn}{A_{\text{FN}}}
\newcommand{\Bfn}{B_{\text{FN}}}



\section{Introduction}
\label{sec:intro}

A typical large area field emitter (LAFE) consists of a multitude of protrusions or emitting sites packed together
in a small area which acts as a source of cold electrons on application of a moderate electric
field \cite{spindt76,teo,dams2012,parmee,basu2015}. These
emitting sites influence each other in a process referred to as shielding. Thus, even when all such emitters
are identical in all respects, the local electric fields at their tips can vary enormously \cite{db_rudra,db_fef,rudra_db_2019,db_rudra_2020,db_hybrid_2020,adc2020,rudra_db_2021}. Sites which
are under relatively weaker influence of neighbouring emitters have
higher local fields and are considered active as they emit electrons readily while others that are more
severely affected by shielding remain passive
until the applied electric field is increased sufficiently for measurable currents to flow. Thus, sparse or non-uniform
emission is more likely to be seen in a glow pattern at lower fields while a more uniform emission
may be observed at higher fields\cite{cole2014,fedoseeva2015,dwivedi2021}.

It has recently been shown numerically that
randomly distributed emitting sites
or random LAFEs, display scaling properties and can therefore be characterized by the emission dimension \cite{rudra_db_2021}
which captures the degree to which the emission is sparse or uniform at a given field strength.
This is significant departure from the usual approach to LAFE characterization which assumes a certain empirical form
of the field emission equation inspired by single emitters. For single curved emitters, the net field
emission current may be expressed as \cite{FN,murphy,jensen2003,forbes2006,FD2007,DF2008,forbes2008,jensen_book}

\be
I = \ca(E_a) J_{\text{MG}}(E_a)  \label{eq:Is}
\ee

\noi
where  $\ca(E_a)$ is a notional emission area and $J_{\text{MG}}(E_a)$ is the Murphy-Good (MG) current density

\be
J_{\text{MG}}(E_a) = \frac{\Afn}{\phi} \text{e}^\eta E_\phi^{\eta/6}  E_a^{2 - \eta/6} e^{-\Bfn \phi^{3/2}/E_a} \label{eq:MG}
\ee

\noi
evaluated at the emitter-apex where the local field $E_a$ is maximum. In the above,
$\eta = \Bfn \phi^{3/2}/E_\phi \approx 9.836~\text{(eV)}^{1/2} \phi^{-1/2}$ where
$E_\phi = (4\pi\epsilon_0/q^3)\phi^2$ is the local field necessary to lower the barrier height
(relative to the Fermi level) to zero, $A_{\text{FN}} = 1.541434 \times 10^{-6}$ $\text{A~eV~V}^{-2}$ and
$B_{\text{FN}} = 6.830890~ \text{eV}^{-3/2} \text{V nm}^{-1}$ are the first and second
Fowler-Nordheim constants, $\phi$ is the local work function of the emitting surface and
$q$ is the magnitude of the electron charge.

For an axially symmetric emitter with a
smooth curved tip and a large aspect ratio, the current $I$ may be
explicitly evaluated by integrating over the
emitter surface\cite{db_dist,rk_gs_db_2021,db_ultram}

\be
I \approx \int J_{\text{MG}}(E(\tth)) 2\pi R_a^2 \frac{\sin \tth}{\cos^4 \tth} d\tth  \label{eq:Iscos}
\ee

\noi
where $\tth$ is a generalized angle defined by

\be
\cos \tth = \frac{z/h}{\sqrt{(z/h)^2 + (\rho/R_a)^2}}
\ee

\noi
with $h$ denoting the height of the emitter, $R_a$ the apex radius of curvature and $z = z(\rho)$
describing the shape of the emitter. The integration in Eq.~(\ref{eq:Iscos}) is possible in
most cases if the apex electric field is known since generic end-caps obey the approximate
relation\cite{db_ultram,physE,db_anode} for the local field $E(\tth) \approx E_a \cos \tth$.
Thus, the notional emission area can be
determined accurately and it is known that it increases approximately
linearly\cite{jensen_book,db_dist,ivnc21} with
the apex electric field (i.e. $\ca \sim E_a$) for sharp tips\cite{db_dist}.
The linear behaviour of $\ca$ together with the expression for the Murphy-Good current density\cite{jensen_book},allows Eq.~(\ref{eq:Is})  to be expressed as\cite{forbes2008,forbes2021}

\be
I = C E_a^{k} e^{-B/E_a}.   \label{eq:Is1}
\ee

\noi
For an axially symmetric single emitter with a generic end-cap, $k = 3 - \eta/6$, $B = \Bfn \phi^{3/2}$ while

\be
C  =  \frac{\ca}{E_a} \frac{\Afn}{\phi} \text{e}^\eta E_\phi^{\eta/6}  \label{eq:As}
\ee

\noi
where $\ca$ is the notional emission area. Note that since $\ca$ increases linearly
with the applied field, $C$ is a constant. If the apex radius of curvature $R_a$ is
known and the end-cap is generic\cite{db_dist,rk_gs_db_2021},
$\ca \approx 2\pi R_a^2 E_a/[\Bfn \phi^{3/2} (1 - f/6)]$ where $f \approx c_s^2 E_a/\phi^2$, 
$c_s$ is the Schottky constant with $c_s^2 \approx 1.44~ \text{eV}^2 \text{V}^{-1} \text{nm}$.

The net emission current in Eq.~(\ref{eq:Is1}) is in fact better expressed in terms of a
measurable quantity such as the applied or macroscopic field $E_0$ or the applied
voltage $V_g$. The apex electric field $E_a$ can thus be written as
$E_a = \gamma_a E_0$ or $E_a = \beta V_g$. The quantity $\gamma_a$ is referred to as
the apex field enhancement factor and for a parallel plate setup $\beta = \gamma_a/D$
where $D$ is the distance between the plates. Thus, for an emitter obeying
Eq.~(\ref{eq:Is1}), a power-k MG plot ($\ln(I/E_0^{k})$ vs $1/E_0$) with $k = 3 - \eta/6$
has a slope equal to $\Bfn \phi^{3/2}/\gamma_a$. Thus, if the work function $\phi$ is known,
the apex field enhancement factor $\gamma_a$ can be extracted from the slope
while the intercept gives $\ln(A \gamma_a^{k})$ which can be used to extract
the notional area $\ca$ or the apex radius of curvature $R_a$.

While Eq.~(\ref{eq:Is1}) and (\ref{eq:As}) serve well to describe electron emission
from a single emitter in the presence of weak or negligible space charge\cite{SCL2020,rk_gs_db_2021,JAP_SC2021},
the situation
becomes somewhat complex for a large area field emitter even from a theoretical
point of view. As mentioned earlier, the presence of a large number of emitters leads to
shielding of field lines and a lowered apex field enhancement factor for each of the
individual emitters. Importantly, the extent to which the apex field is lowered is
not uniform and there may exist a wide variation so that not all emitters
may contribute at a given applied field. The net current can be formally expressed
as a sum over single emitter currents\cite{db_rudra}:

\bea
I_{\text{L}} & = & \sum_{i=1}^{N} I_i = \sum_{i=1}^{N} J_{\text{MG}}^{\text{i}}\ca_i \nonumber \\
& = &  J_{\text{MG}}^{\text{S}} \ca_{\text{S}} \left\{1 + \sum_{i \neq S}  \frac{\ca_i}{\ca_{\text{S}}} \frac{J_{\text{MG}}^{\text{i}}}{J_{\text{MG}}^{\text{S}}} \right\} \nonumber \\
& = & J_{\text{MG}}^{\text{S}} \ca_{\text{S}}  \left\{1  + \sum_{i \neq S}  \frac{\ca_i}{\ca_{\text{S}}}  \left(\frac{\gamma_i}{\gamma_{\text{S}}} \right)^{k} e^{-\frac{B_{\text{S}}}{E_0}(\frac{1}{\gamma_i} - \frac{1}{\gamma_{\text{S}}})} \right\}  \nonumber \\
& = &  J_{\text{MG}}^{\text{S}} \ca_{\text{S}} N_{\text{eff}} = J_{\text{MG}}^{\text{S}} \ca_{\text{L}} = I_{\text{S}} N_{\text{eff}}  \label{eq:Il}
\eea

\noi
where the super(sub)-script $\text{S}$ refers to the super-emitter with the maximum field
enhancement factor, $\ca_{\text{S}}$ is the notional area of the super-emitter,
$\ca_i$ the notional area of the $i^{\text{th}}$ emitter,
and $\ca_{\text{L}}$ is the notional area of the entire LAFE. As such,
the term in the curly bracket in Eq.~(\ref{eq:Il})
should denote the effective number of emitters $N_{\text{eff}}$
that contribute at the level of the super-emitter at a given field.

The formal expression contained in Eq.~(\ref{eq:Il}) is however unsuitable for analyzing
experimental data. For purposes of characterizing a LAFE experimentally, the net
emission current is expressed empirically as\cite{forbes2009}

\be
I_{\text{L}} =  C_{\text{L}} E_0^{k_{\text{L}}} e^{-B_{\text{L}}/E_0},  \label{eq:Il1}
\ee

\noi
a form similar to the single-emitter case with
$B_{\text{L}} = \Bfn \phi^{3/2}/\gamma_{\text{eff}}$ and $C_{\text{L}}$ a constant.
While it is difficult to connect Eq.~(\ref{eq:Il}) and (\ref{eq:Il1}), Fowler-Nordheim (FN) plots
($\ln(I_{\text{L}}/E_0^2)$ vs $1/E_0$) are often approximately linear and the occasionally observed
nonlinearity\cite{forbes2013} may be ascribed to the lack of knowledge about $k_{\text{L}}$ or to the
distribution of emitter enhancement factors. A 2-component model  with two
distinct field enhancement factors and effective emission areas,
has for instance been used to reproduce the non-linearity by adjusting the
parameters appropriately\cite{lu2006,assiss2016}.

The aim of this paper is to take a fresh look at the interpretation of the various
terms in Eq.~(\ref{eq:Il1}) using a model  that is
fairly accurate in determining \{$\gamma_i$\} using the emitter locations on the LAFE.
In particular, we are interested in knowing about the value of $k_{\text{L}}$,
the interpretation of $\gamma_{\text{eff}}$ (or $B_{\text{L}}$),
the nature of $C_{\text{L}}$ and finally
the field dependence of the notional emission area and its relationship with $C_{\text{L}}$
and $B_{\text{L}}$.  The interest in the present work thus
lies beyond the parametric modeling of nonlinearity in an FN plot.

It is clear that $J_{\text{MG}}^{\text{S}}$ contributes $E_0^{2 - \eta/6}$ while $\ca_{\text{S}} \sim E_0$.
Thus, the value of the exponent $k_{\text{L}}$ must be at least $3 - \eta/6$ since the effective number
of emitters is also expected to increase with $E_0$ and can have an additional contribution
to $k_{\text{L}}$ if $N_{\text{eff}}$ has a power law behaviour. In the following, our focus
will on the quantity $N_{\text{eff}}$ and we shall study its variation with the applied
electric field. We shall surprisingly find that $N_{\text{eff}}$ increases much faster than
a power law and that gives rise to the question of interpreting $\gamma_{\text{eff}}$.
In section \ref{sec:method}, we shall describe the model used to arrive at the
field enhancement factors of emitting sites in a random LAFE followed by an evaluation
of $N_{\text{eff}}$ in section \ref{sec:results}.

\section{Obtaining approximate field enhancement factors in a LAFE}
\label{sec:method}

A LAFE typically has tens of thousands of emitting sites placed randomly.
Simulating such a configuration using finite or boundary element methods
is a daunting task even on a parallel machine. An approximate method
is thus useful provided errors are small. The hybrid model provides
a useful and handy alternative. It is based on the (nonlinear) line
charge model\cite{harris15,harris16,db_nonlinear,db_fef} and can be used to determine enhancement factors within
3\% accuracy when the mean inter-pin separation $c$ is greater than or equal to the
height $h$ of individual emitters, increasing to about 8\% when the mean
separation is about $2h/3$. We shall restrict
ourselves to identical hemiellipsoidal emitters and provide an outline of the
main results that can be used to determine the spectrum of
apex field enhancement factors.

Consider a LAFE comprising of $N$ identical hemi-ellipsoidal
emitters, placed at \{($x_i,y_i$)\}, $i = 1,N$. 
The apex enhancement factor $\gamma$ of an $i^{th}$ emitter in the LAFE is given as \cite{db_fef,db_rudra},

\be
\gamma_i  \simeq  \frac{2h/R_a}{\ln\big(4h/R_a\big) - 2 + \alpha_{S_i}}  \label{eq:gamN0}
\ee

\noi
where $\alpha_{S_i}  \approx \sum_{j\ne i}  \alpha_{S_{ij}}$  and

\be
\alpha_{S_{ij}}  =  \frac{1}{\delta_{ij}}\Big[1 - \sqrt{1 + 4\delta_{ij}^2} \Big] + \ln\Big|\sqrt{1 + 4\delta_{ij}^2} + 2\delta_{ij} \Big| \nonumber
\ee

\noi
with $\delta_{ij} = h/\rho_{ij}$, $\rho_{ij} = [(x_i - x_j)^2 + (y_i - y_j)^2]^{1/2}$ being the distance between the $i^{th}$ and $j^{th}$ emitter on the cathode plane. The value of $\alpha_{S_i}$ holds so long as the emitters are not too close and 
under this approximation, the shielding factor, $\alpha_{S_i}$ can be well approximated
by a purely geometric quantity ($ \sum_{j\ne i}  \alpha_{S_{ij}}$ )
which depends only on the relative positions of the emitters.

It is clear that if the emitter locations are randomly distributed, $\alpha_{S_i}$ are distinct.
The presence of anode in close proximity can also be accommodated within
the hybrid model\cite{db_anode,db_rudra_2020,db_hybrid_2020}. In the present case,
we shall consider the anode to be far away without any loss of generality.

The field enhancement factors \{$\gamma_i$\} together with the apex radius of curvature
$R_a$ can be used to determine the total LAFE current, $I_{\text{L}}$ as\cite{db_dist}

\be
I_{\text{L}} \approx \sum_{i=1}^N 2\pi R_a^2 g_i J_{\text{MG}}^i  \label{eq:lafeI}
\ee

\noi
where the area factor $g_i$ is

\be
g_i = \frac{\gamma_i E_0}{B_{\text{FN}} \phi^{3/2}} \frac{1}{(1-f_i/6)}.
\ee

\noi
Eq.~(\ref{eq:lafeI}) can be used to determine the current from a collection of N-emitters
and is well suited for determining the variation of $N_{\text{eff}}$ with $E_0$.

\section{Results}
\label{sec:results}

We shall consider a LAFE with $N = 360000$ identical hemiellipsoidal pins of height $h = 1500\mu$m and
base radius $b = 12.5\mu$m. These are distributed uniformly on a square area with their centres located at
\{($x_i,y_i$)\} and having a mean separation $c$. The side-length of the square LAFE is thus
$\sqrt{N} c$. Eq.~(\ref{eq:gamN0}) allows computation of each enhancement factor. The distribution
of enhancement factors is unimodal as observed in Figs. 1 and 6 of Ref. [\onlinecite{rudra_db_2021}]
for an identical LAFE with $c = 1000\mu$m and $1500\mu$m respectively. The individual enhancement factors can
be used to compute individual currents $I_i$. The sum of these add up to be the total
current $I_{\text{L}}$ while the current of the super-pin having the largest enhancement factor is
denoted by $I_{\text{S}}$. The effective number of superpins at any applied field $E_0$ is
$N_{\text{eff}}(E_0) = I_{\text{L}}(E_0)/I_\text{S}(E_0)$.

\begin{figure}[htbp]
  \vskip -0.75cm
  \hspace*{-0.8cm}\includegraphics[width=.6\textwidth]{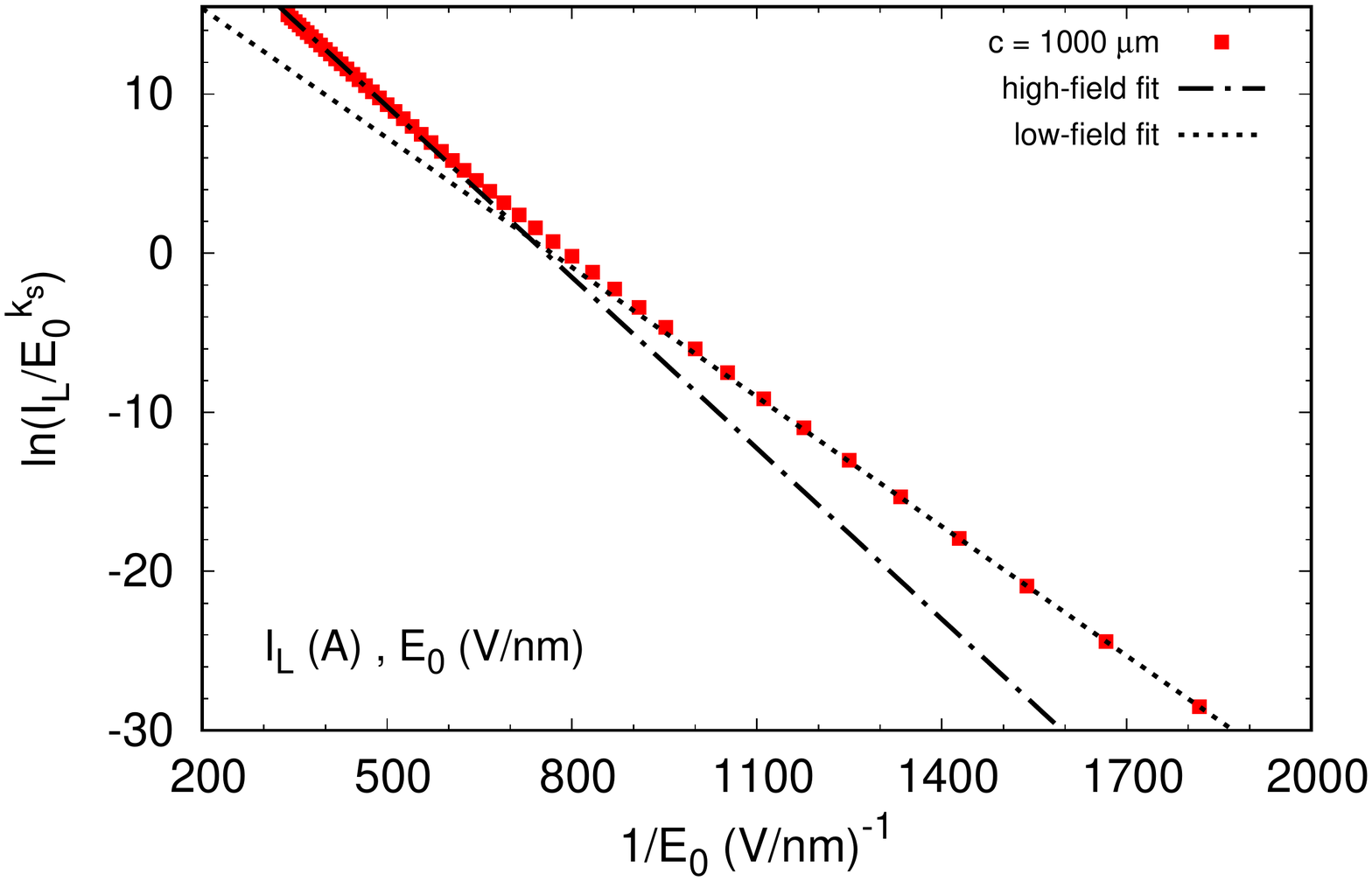}
  \vskip -0.75cm
  \caption{The power-k MG plot shows nonlinear behaviour for $ c = 1000\mu$m and $c/h = 2/3$.
  Here $k_{\text{S}} = 3 - \eta/6$.}
\label{fig:MG_1000}
\end{figure}

\begin{figure}[htbp]
  \vskip -0.75cm
  \hspace*{-0.8cm}\includegraphics[width=.6\textwidth]{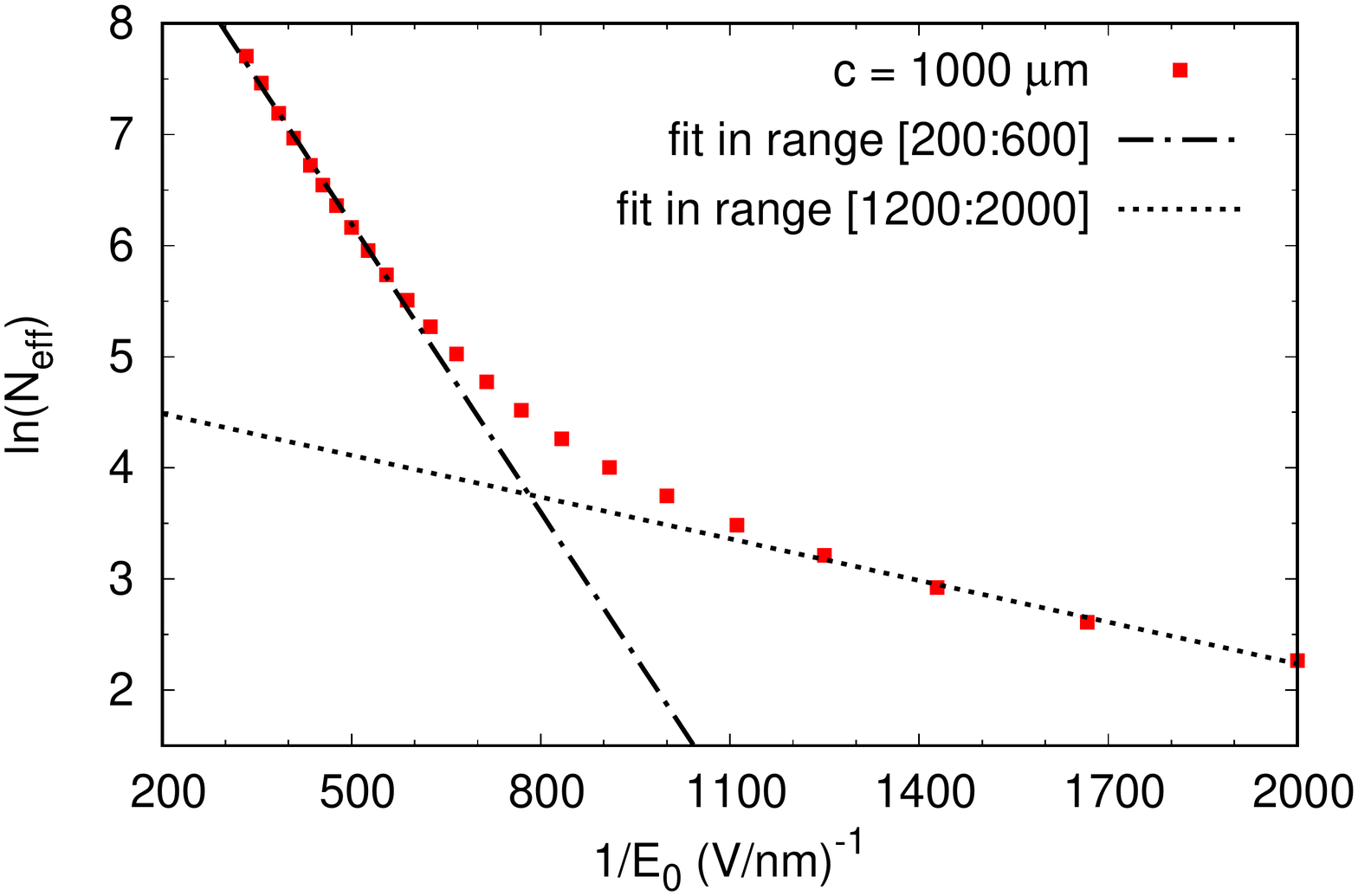}
  \vskip -0.75cm
  \caption{The corresponding plot of $N_{\text{eff}}$ for $c = 1000\mu$m and $c/h = 2/3$.
    The nonlinearity in the power-k MG plot of Fig.~\ref{fig:MG_1000} can be
    ascribed to the nonlinear behaviour of $N_{\text{eff}}$. }
\label{fig:Neff_1000}
\end{figure}

Consider first the case where the mean separation $c = 1000\mu$m. A power-k MG plot is
shown in Fig.~\ref{fig:MG_1000} with $k_{\text{S}} = k = 3 - \eta/6$ being the pre-exponential
exponent of the super-emitter. The plot is clearly nonlinear
and such behaviour is often observed in experiments\cite{lu2006,popov2019,popov2020}.
The corresponding behaviour for $N_{\text{eff}}$ is shown in
Fig.~\ref{fig:Neff_1000}, again plotted against $1/E_0$. The plot has a similar trend
with  a smaller slope at low field and a larger slope at high fields.

Clearly, $N_{\text{eff}}$ can be expressed as

\be
N_{\text{eff}} = N_0 e^{-\alpha/E_0}
\ee

\noi
where $\alpha$ is in general dependent on $E_0$ and expected to vary only weakly as $c$ increases.
The quantity $N_0$ is greater than 1 since $N_{\text{eff}}$ must at least be 1
even at low field. The net LAFE current may therefore be expressed for a LAFE with emitters having smooth end-caps as

\bea
I_{\text{L}} & = & I_{\text{S}} N_{\text{eff}} =  \ca_{\text{S}}  J_{\text{MG}}^{\text{S}} N_{\text{eff}} =  \ca_{\text{L}} J_{\text{MG}}^{\text{S}}  \nonumber \\
& = & \left\{ \frac{2\pi R_a^2}{\Bfn \phi^{3/2}} \frac{\gamma_{\text{S}} E_0}{1 - f_{\text{S}}/6}  N_0 e^{-\alpha/E_0}\right\}  J_{\text{MG}}^{\text{S}} \nonumber \\
& = & C_{\text{L}} E_0^{k_{\text{L}}} e^{-B_{\text{L}}/E_0} \label{eq:Il2}
\eea

\noi
where

\bea
C_{\text{L}}& =& \frac{2\pi R_a^2 N_0}{\Bfn \phi^{3/2}} \frac{\Afn}{\phi} e^\eta E_\phi^{\eta/6} \frac{\gamma_{\text{S}}^{k_{\text{S}}}}{1 - f_{\text{S}}/6} \\
k_{\text{L}} & = & 3 - \eta/6 \\
B_{\text{L}}  & = & \frac{\Bfn \phi^{3/2}}{\gamma_{\text{S}}} + \alpha(E_0)
\eea

\noi
while, the notional area $\ca_{\text{L}}$ is

\be
\ca_{\text{L}} = \frac{2\pi R_a^2 N_0}{\Bfn \phi^{3/2}} \frac{\gamma_{\text{S}} E_0}{1 - f_{\text{S}}/6}  e^{-\alpha/E_0}. \label{eq:notional}
\ee

\noi
Note that even though $f_{\text{S}}  = c_s^2 (\gamma_{\text{S}} E_0)/\phi^2$ depends on the applied field $E_0$, $f_{\text{S}}/6$ is
a small quantity compared to unity, so that the dominant pre-exponential field dependence in $I_{\text{L}}$
is $E_0^{3 - \eta/6} = E_0^{k_{\text{S}}}$.

A plot of $\ln(I_{\text{L}}/E_0^{k_{\text{S}}})$ vs $1/E_0$ must have a (local) slope $-\cal{S}$
and intercept ${\cal{I}}$ with 

\bea
    {\cal{S}} & = & \left( \frac{\Bfn \phi^{3/2}}{\gamma_{\text{S}}} + \alpha \right) + \frac{1}{E_0} \frac{d\alpha(E_0)}{dE_0^{-1}~~} \label{eq:slope}\\
    {\cal{I}} & = & \ln C_{\text{L}} +  \frac{1}{E_0} \frac{d\alpha(E_0)}{dE_0^{-1}~~}  \label{eq:intercept}
\eea

\noi
It follows from Eq.~(\ref{eq:slope}) that

\be
\gamma_{\text{S}} = \frac{\Bfn \phi^{3/2}}{{\cal{S}} - \alpha - \alpha'/E_0 } \label{eq:gammaM}
\ee

\noi
where $\alpha' = \frac{d\alpha(E_0)}{dE_0^{-1}~~}$. Thus, $\gamma_{\text{S}}$ is related to the slope ${\cal{S}}$,
$\alpha$ and $\alpha'$ and hence cannot be accurately extracted from the slope and intercept. In fact there are more unknowns
($\alpha,\alpha',\gamma_{\text{S}}$ and $R_a^2 N_0$) than the number of measurable quantities and
hence extraction of LAFE characteristics such as $\gamma_{\text{S}}$ and the notional area poses a challenge.

\begin{figure}[htbp]
  \vskip -0.75cm
  \hspace*{-0.8cm}\includegraphics[width=.6\textwidth]{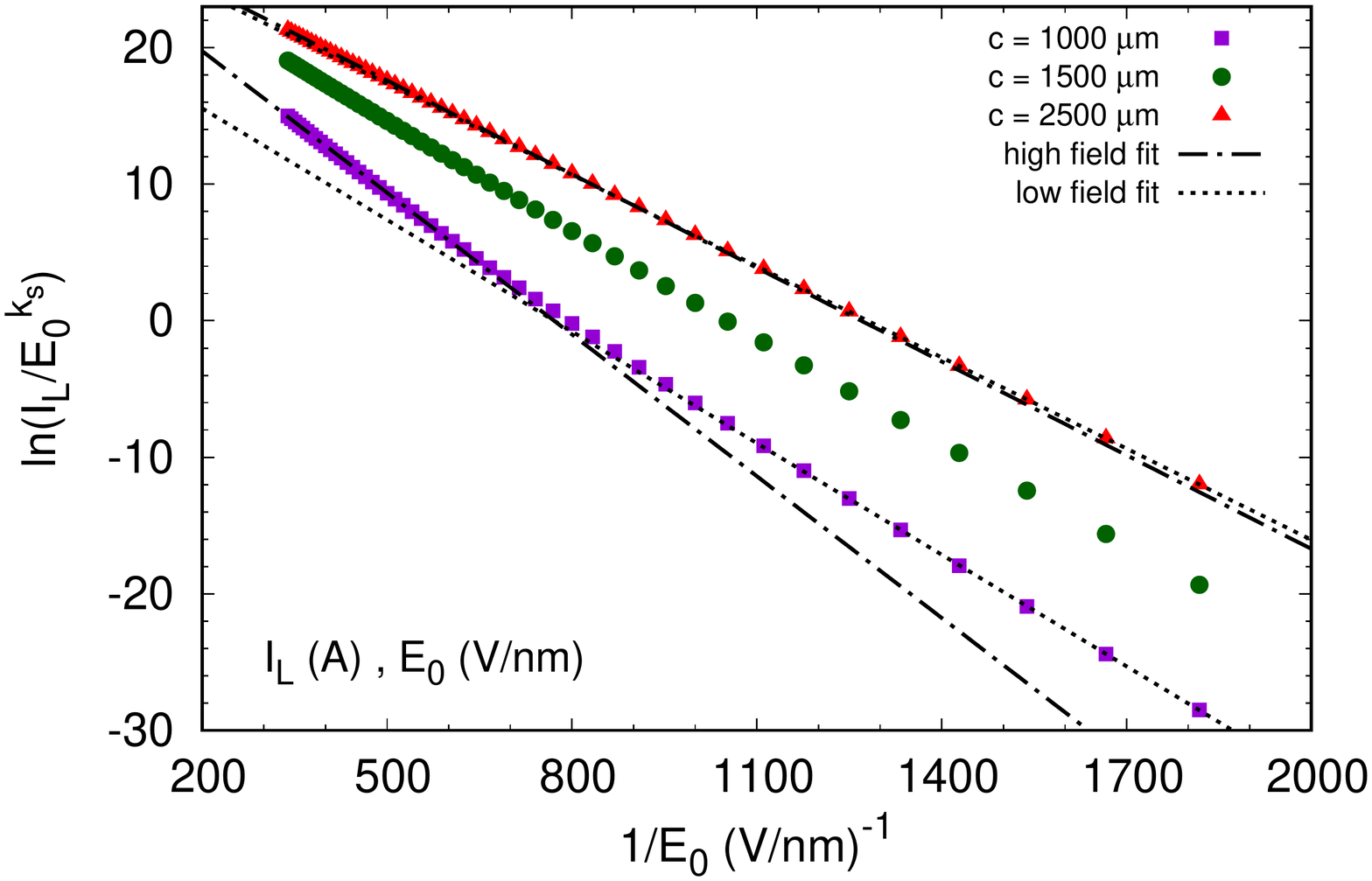}
  \vskip -0.75cm
  \caption{The power-k MG  plot with $k_{\text{S}} = 3 - \eta/6$ shows nonlinear
    behaviour for $ c = 1000\mu$m but is nearly linear for $c = 2500\mu$m.
    Also shown are the low-field and high-field straight line fits. The
    difference in slope is negligible for $c = 2500\mu$m.
  }
\label{fig:MGcompare}
\end{figure}

Before we proceed to discuss the extraction of approximate LAFE characteristics,
it is important to explore whether such a behaviour is limited to a dense LAFE
such as at $c = 1000\mu$m or exists in general. Fig.~\ref{fig:MGcompare} shows a comparison for three
different mean spacings, $c = 1000\mu$m, $1500\mu$m and $2500\mu$m. While, $c = 1000\mu$m
displays nonlinearity, $c = 2500\mu$m is nearly linear as testified by the closeness of the
high and low field fits.

\begin{figure}[htbp]
  \vskip -0.75cm
  \hspace*{-0.8cm}\includegraphics[width=.6\textwidth]{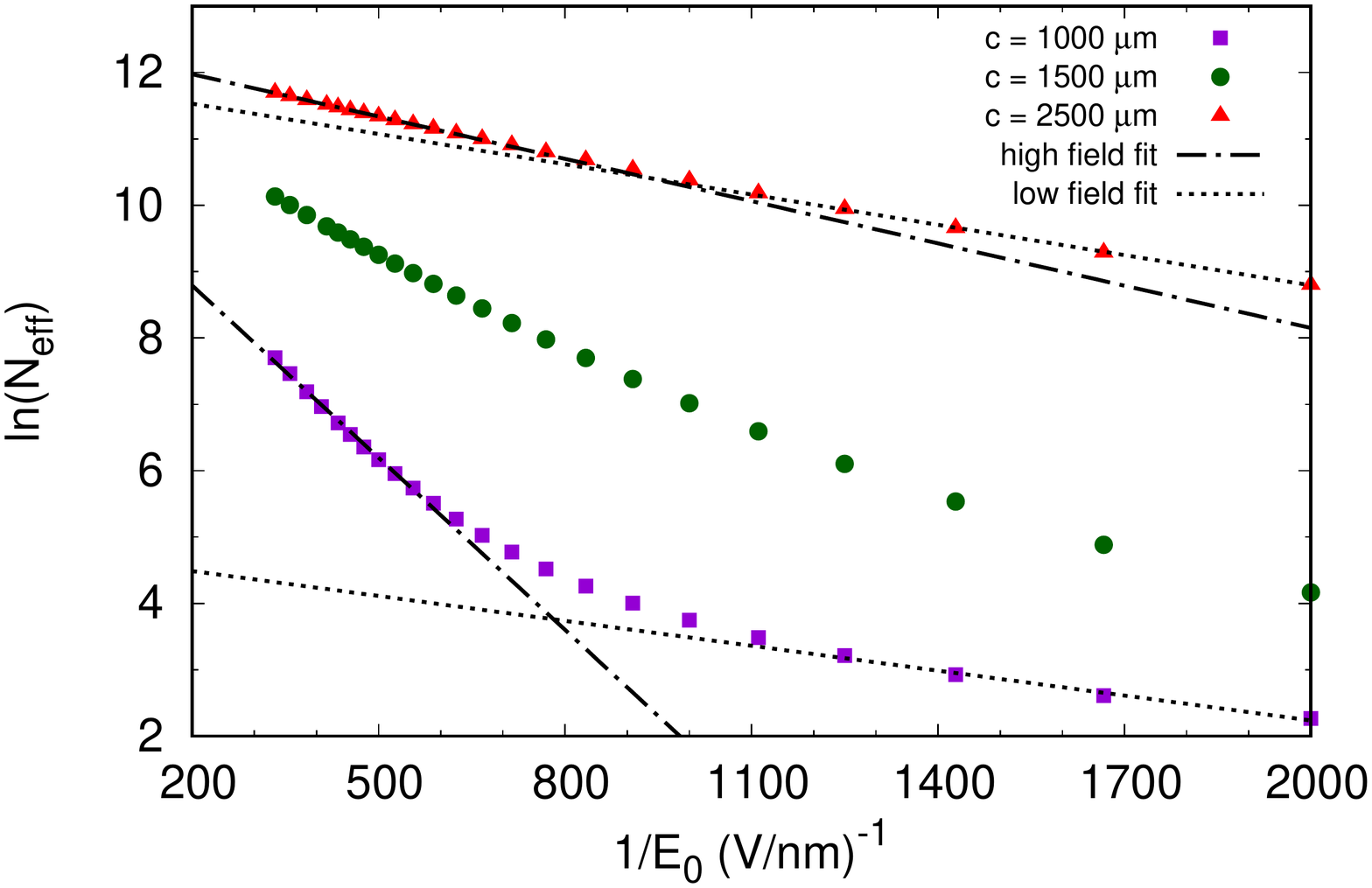}
  \vskip -0.75cm
  \caption{The corresponding plot of $N_{\text{eff}}$ for $c = 1000\mu$m, $1500\mu$m
    and $2500\mu$m with $h = 1500\mu$m.
    }
\label{fig:Neffcompare}
\end{figure}

The corresponding behaviour of the $N_{\text{eff}}$ is shown in Fig.~\ref{fig:Neffcompare}.
The nonlinearity is large for $c = 1000\mu$m but decreases considerably at
$c = 2500\mu$m. Clearly the exponential behaviour persists at $c = 2500\mu$m as evident
from the straight line fits signalling a non-zero value of $\alpha$.
Note however that the overall slope ($\alpha$) decreases at larger spacing as
expected since the large $c$ limit must correspond to $\alpha \rightarrow 0$
since the emitters behave independently as $N$ single emitters, each having
the same enhancement factor.

The exponential dependence of $N_{\text{eff}}$ on $E_0^{-1}$ makes interpretation of
the slope and intercept of an power-k MG plot, a nontrivial matter. Traditionally, the slope
has been used to extract $\gamma_{\text{eff}}$ or $\gamma_{\text{S}}$ while
the intercept has been linked to the notional emission area and the area efficiency
of a LAFE. It is clear from Eq.~(\ref{eq:gammaM}) and (\ref{eq:notional}) that
$\alpha$ and its derivative $\alpha'$ with respect to $E_0^{-1}$,
plays a crucial role in the  determination of $\gamma_\text{S}$ and $\ca_{\text{L}}$.

Since, $\alpha$ and $\alpha'$ are small at very low fields (large $E_0^{-1}$), an approximate evaluation
of $\gamma_\text{S}$ and $C_{\text{L}}$ can be made by setting $\alpha = 0$ and $\alpha' = 0$ in Eq.~(\ref{eq:gammaM}). Thus,

\bea
\gamma_\text{S} & \approx & \frac{\Bfn \phi^{3/2}}{{{\cal{S}}_{lo}}} \label{eq:appgam} \\
\ln & ~C_{\text{L}}  \approx & {{\cal{I}}_{lo}}. \label{eq:appC}
\eea

\noi
In the above, the subscript `$lo$' refers to the low field region. The value of $\ln C_{\text{L}}$
can in turn be used to estimate $R_a^2 N_0$. Having extracted approximate
values of $\gamma_\text{S}$ and $R_a^2 N_0$ from the low field region, the value of $\alpha(E_0)$ at any higher
range of field can be extracted as

\be
\alpha(E_0) = {\cal{S}}(E_0) - {\cal{S}}_{lo} - E_0 \left[ {\cal{I}}(E_0) - {\cal{I}}_{lo} \right] \label{eq:appalp}
\ee

The estimates of $\gamma_\text{S}$, $\ln C_{\text{L}}$ (hence $R_a^2 N_0$) and $\alpha(E_0)$ determined
through Eqns. (\ref{eq:appgam}), (\ref{eq:appC}), and (\ref{eq:appalp}) are only approximate. They
can be in turn be used to determine the notional area $\ca_{\text{L}}(E_0)$
using Eq.~(\ref{eq:notional}). The initial errors in determining $\gamma_\text{S}$ and $\ln C_{\text{L}}$
will no doubt propagate in evaluating $\alpha(E_0)$ and hence $\ca_{\text{L}}(E_0)$. The notional
area is thus prone to larger errors due to the exponential dependence on $\alpha(E_0)/E_0$.
For the systems considered here, $\gamma_{\text{S}}$ can
be determined using this procedure to within 4-7\% error using a linear fit in the low field region.

\section{Summary and Conclusions}

The purpose of this paper has been to highlight some of the pitfalls in analyzing
and characterizing a large area field emitter. To this end, we have considered an ideal
LAFE consisting of 360000 identical hemiellipsoidal emitters with smooth endcaps, and used
an approximate model to determine the individual apex field enhancement factors.
The total LAFE current was then determined by summing over the current from each emitter
and subsequently expressed as the product of current from a super-emitter and the
effective number of super-emitters ($N_{\text{eff}}$) at any applied field. On analyzing the data,
it was found that $N_{\text{eff}}$ increases as $N_0 e^{-\alpha/E_0}$, thereby contributing
an additional term to the slope of a power-k  MG plot and posing difficulties in its interpretation.
An approximate method for extraction of the apex field enhancement factor of the
super-emitter was suggested using the low field limit where $\alpha$ may be neglected.
The notional area and its increase with applied field can then be estimated using the
slope and intercept at higher field values.

One of the most striking features of this study has been the behaviour of the
notional emission area with applied field. The exponential dependence
$\ca_{\text{L}} \sim E_0 e^{-\alpha/E_0}$ stands in sharp contrast to the common
assumption that the area depends only weakly on the applied field and
may be ignored as a first approximation for a LAFE. This sharp increase
in the notional area with the applied field, makes a LAFE distinct from a
single emitter where the increase is linear for generic smooth end-caps.

The distribution of field enhancement factors gives rise to the
nonlinearity in power-k MG plot especially when the mean separation is
less than the height of the emitters. For identical emitters and large mean
separations, it is expected that all emitters will have nearly identical
enhancement factors so that a linear behaviour should ensue. In that case,
the field enhancement factor can be directly extracted from the slope of the power-k MG plot.
If, however, the emitters have a distribution in height
or apex radius of curvature, there will be a distribution of enhancement
factors even when the mean separation is large. Thus, the slope of the
$\ln N_{\text{eff}}$ plot need not fall off to zero at larger separation and the power-k MG plot may continue
to display nonlinearity. The predictions for the approximate field enhancement factor of the super-emitter
or the notional area, using Eqns. (\ref{eq:appgam}), (\ref{eq:appC}), and (\ref{eq:appalp}), will
have larger errors even for this simplest non-ideal case.
At the moment, it is hard to predict the behaviour of
graphene or carbon-film based LAFEs but it is likely that the change in active
emitting sites with applied field makes the connection between the slope of a power-k
MG plot and the effective enhancement factor somewhat dubious.

Finally, the question of orthodoxy\cite{forbes2013,kolosko2016} in field emission may need to be revisited
since even for an ideal LAFE, the scaled barrier field is not directly related to the slope of
an IV curve.

\section{Acknowledgements} The author thanks Rashbihari Rudra,
Raghwendra Kumar, Rajasree R. and Gaurav Singh for discussions, suggestions and
a critical reading of the manuscript.

\section{Author Declarations}

\subsection{Conflict of interest} There is no conflict of interest to disclose.
\subsection{Data Availability} The data that supports the findings of this study are available within the article.

\vskip 0.05 in

\end{document}